\documentclass{egpubl}
\CGFStandardLicense

\usepackage[T1]{fontenc}
\usepackage{dfadobe}  

\usepackage{cite}  
\BibtexOrBiblatex
\electronicVersion
\PrintedOrElectronic
\ifpdf \usepackage[pdftex]{graphicx} \pdfcompresslevel=9
\else \usepackage[dvips]{graphicx} \fi

\graphicspath{{figs/}{figures/}{pictures/}{images/}{./}}
\usepackage{tabu}                      
\usepackage{booktabs}                  
\usepackage{lipsum}                    
\usepackage{mwe}                       

\usepackage{mathptmx}                  

\usepackage{comment}
\usepackage{diagbox}
\usepackage{tabu}
\usepackage{algorithm}
\usepackage{algpseudocode}
\usepackage{amsmath}
\usepackage{cancel}
\usepackage{cleveref}

\usepackage{siunitx}
\usepackage{dcolumn}

\usepackage{egweblnk} 

\title[Inverse Garment and Pattern Modeling with a Differentiable Simulator]%
      {Inverse Garment and Pattern Modeling with a Differentiable Simulator}

\author[Boyang Yu \& Frederic Cordier \& Hyewon Seo]
{\parbox{\textwidth}{\centering Boyang Yu $^{1}$\orcid{0000-0002-0934-610X} \, Frederic Cordier $^{2}$\orcid{0000-0003-0675-5431}
        and Hyewon Seo$^{1}$\orcid{0000-0001-8851-0256} 
        }    
        \\
{\parbox{\textwidth}{\centering $^1$ ICube laboratory, CNRS–University of
Strasbourg, France \\
         $^2$ IRIMAS, University of Haute-Alsace, France%
       }
}
}

\begin{document}


\maketitle
\begin{abstract}
The capability to generate simulation-ready garment models from 3D shapes of clothed people will significantly enhance the interpretability of captured geometry of real garments, as well as their faithful reproduction in the digital world. This will have notable impact on fields like shape capture in social VR, and virtual try-on in the fashion industry. To align with the garment modeling process standardized by the fashion industry and cloth simulation software, it is required to recover 2D patterns, which are then placed around the wearer's body model and seamed prior to the draping simulation. This involves an inverse garment design problem, which is the focus of our work here: Starting with an arbitrary target garment geometry, our system estimates its animatable replica along with its corresponding 2D pattern. Built upon a differentiable cloth simulator, it runs an optimization process that is directed towards minimizing the deviation of the simulated garment shape from the target geometry, while maintaining desirable properties such as left-to-right symmetry. Experimental results on various real-world and synthetic data show that our method outperforms state-of-the-art methods in producing both high-quality garment models and accurate 2D patterns.

\printccsdesc   
\end{abstract}  
\section{Introduction}
The ability to generate simulation-ready garment digital twins from 3D shapes of dressed people has a wide range of applications in virtual try-on, garment reverse engineering, and social AR/VR. It will allow, from the retrieved garment models, to obtain new animation, or to better capture and interpret subsequent garment geometry undergoing deformation. This is particularly compelling given the increasing accessibility of detailed 3D scans of people with clothing. Such a garment recovery system should ideally satisfy the following: high fidelity to faithfully replicate the given 3D geometry, adaptability to obtain new garment simulations on different body shapes and poses, and the ability to recover 2D patterns, to conform to the standard garment modeling processes used in both the fashion industry and cloth simulation software.

In this paper, we address the challenging problem of converting a given 3D shape of a dressed garment to an animatable form by estimating its precise 2D pattern shape. 
Such pattern-based modeling closely mimics the design process for both real-world and synthetic garments, and effectively disentangles the inherent shape from deformations caused by external forces and internal fabric properties during draping. Based on a differentiable physics-based simulator, our system solves an inverse simulation problem: iteratively optimizing both the pattern shape and physical parameters to ensure that the draped garment mesh on the estimated body aligns with the target garment shape. The ability to estimate garment patterns facilitates the adaptation of the reconstructed garment to new conditions for downstream applications. 
New animations on different body shapes or poses can be synthesized by placing and seaming the produced pattern around the body mesh prior to the draping simulation. In addition, our approach does not require training data and is capable of faithfully replicating intricate garment shapes. We evaluate our approach across various garment types and demonstrate that our method produces patterns and its garment counterparts of promising quality. Compared to the state-of-the-art methods, ours achieves superior performance in terms of both reconstruction and pattern accuracy.

In summary, our contributions are as follows:
\begin{itemize}
    \item A new formulation of inverse pattern modeling based on a physics-based differentiable simulator, producing output that aligns with the current garment modeling and fabrication process.
    \item A compact parameterization of the sewing pattern to streamline the optimization, ensuring the preservation of desirable pattern properties;
    \item An upgraded   differentiable simulator \cite{liang19, narain2012adaptive} aimed at improving computation speed, operating seamlessly within an end-to-end process.
\end{itemize}
Our generated data and code will be made available for research purposes at \url{https://github.com/MLMS-CG/IGPM}.

\begin{figure*}[ht]
\centering
\includegraphics[width=1.0\linewidth]{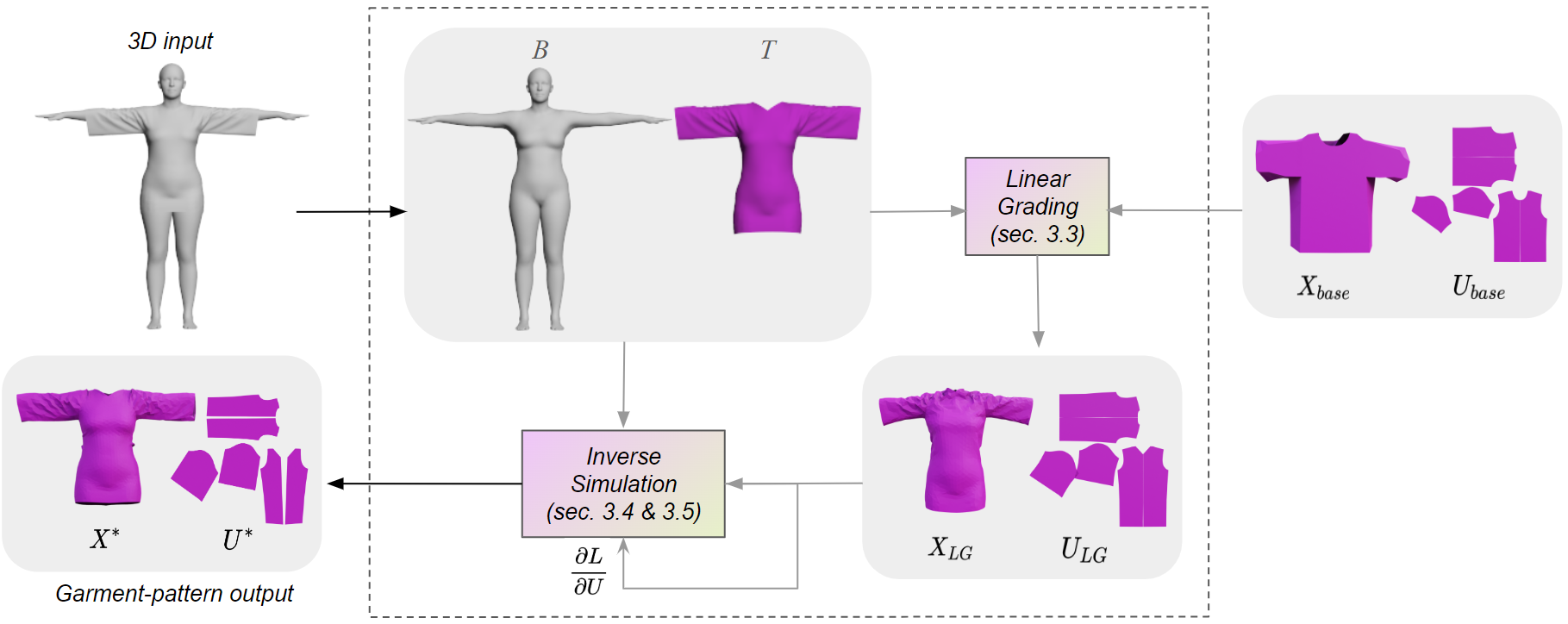}
\caption{Given a 3D shape of a dressed garment, our method generates a simulation-ready garment by estimating its precise 2D pattern shape. The garment mesh T is used as the target in two-phase garment fitting process, where the sewing pattern and material parameters are jointly optimized. First, a linear grading captures the overall proportions and sizes geometrically. This is followed by an inverse simulation, where optimization coupled with a differentiable simulation refine the fit and other geometric details.
}
\label{fig:pipeline}
\end{figure*}

\section{Related work}
\label{sec:related}
\subsection{Garment shape recovery from 2D/3D data}
Current methods, such as NeRF \cite{wang2021nerf, H-Nerfs21} or PIFu \cite{pifu19} reconstruct 3D mesh of a dressed person from one or more 2D images. However, the generated watertight meshes tend to be coarse and can not be directly draped or simulated on a body, even after a 3D segmentation to separate the garment part from the body\cite{bang21, pons2017clothcap, tiwari2020sizer, Chen20TightCap, ma2020cape}. For the reconstructed garment geometry to be reusable to a new wearer, the 2D pattern structure or canonical shape should be recovered to ensure the compatability with existing physics-based or neural simulators. This presents a notable challenge, which we aim to tackle in this work and has been addressed by a few. For example, in \cite{yang18PIgarment}, search-based optimization is utilized to recover both 2D sewing patterns (parameterized by numerical values such as sleeve length, waist width) and the 3D garment, so that it minimizes the 2D silhouette difference between the projected 3D garment and the input image. The draped 3D garment shape is obtained by using a physics-based simulator \cite{narain2012adaptive}. Unlike theirs, in our work the estimation of pattern parameters are coupled with the physically-based simulation process, allowing for direct and precise mapping of 3D garment error to both the pattern geometry and material parameters.

Generative models have demonstrated their fitting capacity to given 2D or 3D inputs. Typically, they solve for optimal parameters in the model to obtain the best estimation of the body and the cloth on the input data. Early models \cite{ patel20tailornet, ma2020cape, mgn2019} learned the 3D clothing deformation as a displacement function, often conditioned on shape, pose, and style of garment deduced from desired or given targets. Such displacement-based representation assumes one-to-one mapping between the body and the garment surfaces, primarily suited for tight-fit clothes that closely conform to the body in terms of both topology and shape. To cope with loose garments, the blend shapes and skinning weights\cite{santesteban2021self} are diffused for 3D points around the body surface, have been used to 
drive the shape of garments according to the body pose. Alternative representations have been explored, such as implicit shape models \cite{corona2021smplicit, buffet2019implicit, santesteban2021ulnefs}, patch-based surface representation \cite{ma2021scale}, or articulated dense point cloud \cite{ma2021pop}.

\subsection{Sewing pattern estimation from 3D data}
Closely related to our work, several recent studies have tackled the challenge of estimating 2D sewing patterns from 3D garment meshes.

\noindent \textbf{3D-to-2D surface flattening.} 
Several works considered the garment mesh as a developable surface \cite{stein2018developability}, obtaining 2D pattern panels by cutting the garment surface into 3D patches and then flattening each patch onto a plane. The cutting lines are found either by projecting the predefined seam lines from the body mesh onto the garment mesh \cite{bang21}, along curvature directions 
\cite{vaxman2016directional, pietroni2022computational} on the surface, or through variational surface cutting\cite{sharp2018variational} to minimize the distortion induced by cutting and flattening \cite{wolff20213d}. While intuitive and versatile, such geometric strategy is prone to generating patterns that deviate from traditional panel semantics, or lack of symmetry, making them unsuitable for garment production. Moreover, purely geometric methods \cite{bang21} do not account for the fabric's elasticity in the physical body-cloth interaction during the draping process, often leading a sewing pattern that cannot accurately replicate the originally designed garment.

\noindent\textbf{Pattern geometry optimization.}
Several works regard a 3D garment as the simulation result of its 2D pattern input, which is similar in spirit to our approach. \cite{bartle16} introduced a two-phase approach comprising direct 3D garment editing and pattern alternation for fitting: a 3D garment shape optimization phase and an inverse 2D pattern design phase. While sharing the main idea of computing the sewing pattern inversely from the garment, they use a quasi-static simulation as a black box. Moreover, the solution obtained in the second phase generally does not meet the result elaborated in the first phase. \cite{wang18} tries to solve only one constrained optimization for adjusting a standard sewing pattern, with a focus on garment grading tasks (e.g., resizing from S to XL) to achieve a good fit on a different body shape. The cost function is designed to accommodate garment grading for various body sizes, ensuring that the resulting garments replicate the same degree of looseness, surface smoothness, and stretching ratio as observed in the base garment on the base body. \cite{wolff20213d} solves another problem of garment customization, by optimizing the garment shape, thereby improving the comfort of the cloth to a specific individual. The  2D pattern is generated by computing patch lines followed by geometric flattening.

\noindent \textbf{Learning based estimation.}
The work of \cite{korosteleva2022neuraltailor} explores a learning-based approach for estimating the sewing pattern of a given 3D garment shape. Leveraging a dataset of 3D garments with known sewing patterns across variety of garment design, their model is capable of regressing the sewing patterns representing the garment, as well as the stitching information among them. While intriguing, their model is limited to the settings presented in the training dataset: Specifically, the model struggles when handling garments with different material properties than those used to generate the dataset. Likewise, it tends to show limited performance on garments draped on bodies other than the average SMPL female body in a T-pose. A plethora of works have followed this approach such as \cite{chen2024panelformer, nag2023personaltailor,he2024dresscode}. In particular, SewFormer\cite{liu23sewformer} trains a two-level Transformer network to regress sewing patterns from a single 2D human image, by using a dataset of images and sewing-pattern pairs covering a wide range of body poses and shapes, as well as garment types. All these learning methods, however, require a substantial dataset, which is expensive to build.

\subsection{Differentiable cloth simulator}
Despite the significant progress in differentiable simulation\cite{liang19, li2022diffcloth, larionov2022estimating, guo2021inversesim}, no previous works have addressed the simultaneous recovery of sewing patterns and the reconstruction of garments from the inputs of clothed humans. Concurrent to our work is DiffAvatar\cite{li2023diffavatar}, who proposes a similar underlying idea of optimizing the sewing pattern and physical parameters to reproduce the captured 3D scan. Among the several distinctive differences between our work and theirs, one notable aspect is that our approach performs a first-stage geometric approximation to capture the overall shape of the target garment, prior to optimization through an inverse simulation. Consequently, our system is less sensitive to the initial pattern. Additionally, unlike theirs, our approach preserves desirable pattern properties such as symmetry.

\subsection{Neural garment draping and dynamic modeling}
Neural garment models train deep neural networks to learn draping or dynamic garment behavior on the wearer's body, aiming to replace computation-intensive physics-based simulations. While early models are trained in a supervised manner over 3D garment dataset
\cite{gnn_vto, bertiche2020pbns, zhang2022motionguided, santesteban2019vto}, later models \cite{bertiche2022neuralcloth, santesteban2022snug} adopt a self-supervised approach with physics-based losses. The idea of self-supervised learning has propelled advances of recent garment models. \cite{grigorev2022hood} combines the graph neural network training with the self-supervision loss terms; \cite{de2023drapenet} conditions on a latent code for more generic garments draping using UDF to represent garment surfaces; \cite{li2024isp} consists of flat 2D panels defined by 2D SDF, and each panel is associated a 3D surface parameterized by the 2D panel coordinates. 
\section{Method}
\label{sec:method}

\subsection{Overview}
In this section, we describe our method outlined in \Cref{fig:pipeline}. Drawing an analogy to garment production, the garment geometry in our work is determined by the style and size of its sewing pattern, which is parameterized for efficient modification 
(\cref{sec:representation}). The first component of our system is the linear grading which accounts for capturing the coarse geometry such as size and proportion (\cref{sec:plg}). The second component further refines the model to capture the detailed garment shape and precise pattern. At the heart of our technique is an optimization-driven pattern refinement based on a differentiable cloth simulator (\cref{sec:sim}, \cref{sec:opt}), where the simulated garment is iteratively altered along with the physical parameters.
\subsection{Representation of base pattern and body}
\label{sec:representation}
\begin{figure}[htp]
\centering
\includegraphics[width=1.0\linewidth]{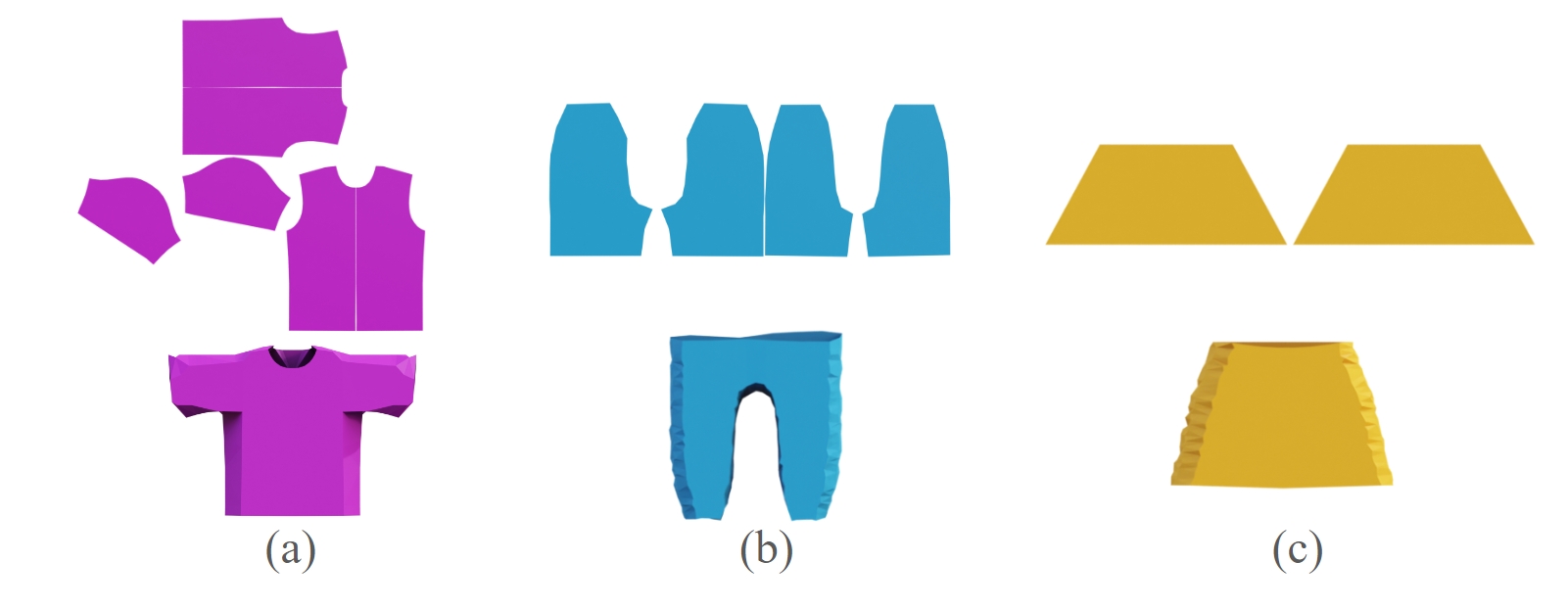}
\caption{Base models for three garment topologies. (a) t-shirt/dress; (b) pants; and (c) skirt.}
\label{fig:bases}
\end{figure}

The garment shape is determined by the shape and size of its sewing pattern, which is a collection of 2D panels that are placed around the wearer's body and stitched together at an initial stage of the later simulation. 
We observe that many garments share a same pattern topology, with geometric variations. Therefore, our system provides several base models selected from the Berkeley Garment Library \cite{narain2012adaptive}. These base models, i.e. 2D patterns and their corresponding sewn 3D meshes ($U_{base}$ and $X_{base}$ in \Cref{fig:pipeline}, respectively), are available for representative garment categories. Although  several methods exist for estimating a base model given a garment mesh, which we could here, we currently let the user select a base model depending on the target garment type. 
Optionally, users can incorporate customized  base models into the system.
\Cref{fig:bases} shows three base models (t-shirt/dress, pants, and skirt) used in our experiments. 

\begin{figure}[ht]
\centering
\includegraphics[width=0.7\linewidth]{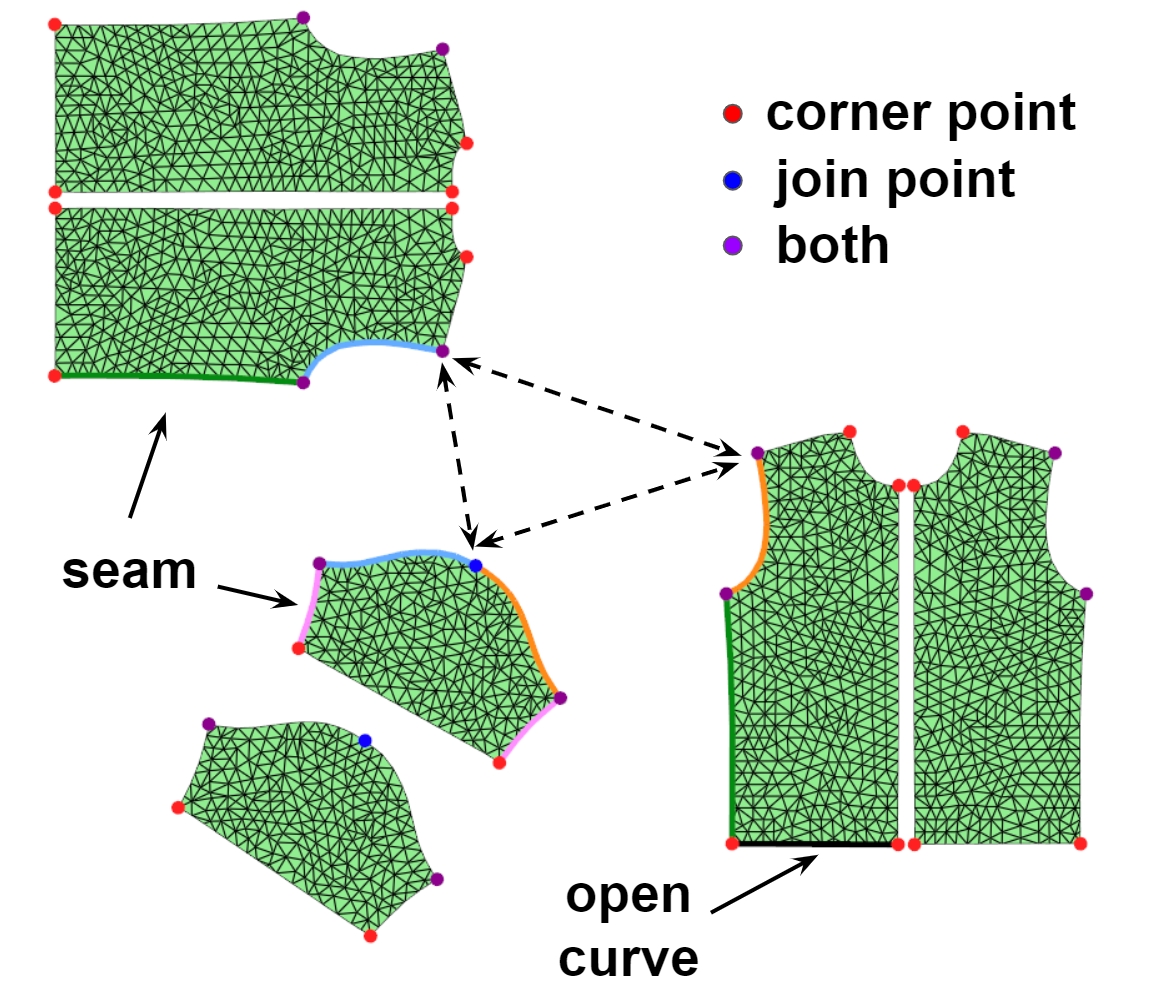}
\caption{
Control points on the T-shirt base pattern mesh. Specific seams are colored consistently across different panels for clarity. }
\label{fig:controlpoints}
\end{figure}

\noindent\textbf{Parameterization.} 
The planar pattern mesh $U$ serves as the FEM reference (prior to any deformation) for its corresponding 3D garment mesh $X$ during the later simulation. The mapping between a 2D vertex $u\in U$ to a 3D node $x \in X$ is known from the base models, in the form of a UV Map. A panel is a 2D triangular mesh bounded by a number of piece-wise curves parameterized by a set of control points. As shown in \Cref{fig:controlpoints}, two curves join at a control point $c^{i}$, which is typically the vertex of $C^0$ curvature discontinuity (\textit{corner point}), or the vertex having two or more \textit{seam-counterparts} belonging to other panels (\textit{join point}) which will be merged into one node at the time of sewing. Note that the same control point can be both a corner and a join point (purple points in \cref{fig:controlpoints}).

The control points are grouped into disjoint sets 
${C}=\{C_{p}\}$, one for each panel $p$. Within a panel, the control points are ordered in a counterclockwise manner, i.e. $C_{p}=\{c_{p}^{i}\}$. 
Throughout the pattern optimization process, the control points serve as variables, while all other mesh points are repositioned to maintain the relative locations with respect to them (See \Cref{sec:plg} for the repositioning method). This effectively reduces the dimensionality of the solution space.

\noindent\textbf{Symmetry detection.} 
Let ${K}$ denote the effective control point group, initially set to $\{C_{p}\}$. To further reduce the dimensionality of ${K}$ and to preserve the pattern symmetry (which is often a desirable property in garment production) during optimization, we detect the pattern symmetry in two steps: 
It first detects \textit{inter-panel symmetry} by computing for each pair of panels an aligning rigid transformation \cite{procrustes1966} and evaluating the quality of alignment. If their alignment score is sufficiently high, we remove one of the two panels from ${K}$. Next, we perform \textit{intra-panel symmetry} detection within ${K}$, by computing for each pair of control points $c_{p}^{i},  c_{p}^{j}\in C_{p}$ its axis of symmetry and evaluating the symmetry score for the remainder of control points in the panel. The control points pair with the highest score exceeding a predefined threshold is used to identify left-to-right symmetry within a panel, subsequently leading to a further reduction of control points from ${K}$. During the symmetry detection, we identify a transformation matrix (flipping, rotation) for each symmetry pair of control points. With the new coordinates of effective control points $K$ obtained from pattern alteration, the full coordinates of set $U$ can be restored via a series of matrix multiplication. The transformation matrices are precomputed once and reused throughout the optimization. The algorithmic description of the symmetry detection is provided in Algorithm 1 of Supplementary material.

\noindent\textbf{Sewn garment shape.}
The sewn 3D garment mesh is made of 2D pattern placed in 3D, topologically stitched along seams, and geometrically deformed to have sufficiently large inter-panel distances in order to avoid any potential body-garment interpenetration. Note that the vertices along the seams will be merged with their seam counterparts on other panel(s) during stitching. Hence, the correspondence between 3D nodes $x \in X$ and 2D vertices $u\in U$ is one-to-many for those on the seams, while it remains one-to-one for the rest. 

\noindent\textbf{Body model.} The draped shape of a garment is determined by not only the pattern shape $U$, but also the underlying body $B$, and their interaction during contact. We adopt the parametric SMPL model \cite{loper2015smpl} to represent the body, for which several efficient registration methods to 3D data exist. We used the method by  \cite{bhatnagar2020ipnet, bhatnagar2020loopreg} to fit SMPL parameters to the wearer's body. The resulting model is denoted as $B=SMPL(\beta^*, \theta^*)$, $\beta$ and $\theta$ are respectively the shape and pose parameters in SMPL.

\subsection{Pattern linear grading}
\label{sec:plg}
In this phase, we aim to perform a preliminary geometric deformation at the panel level to capture the overall geometry of the target garment, such as length and proportion. The main idea is to match closely corresponding 3D open contours on both garments by deriving the relocation of their corresponding 2D open curves (See \Cref{fig:controlpoints} for an example). A 3D open contour is composed of edges connected to only one adjacent triangle, which often carry design features, representing elements such as necklines, hem contours, cuff contours, etc. Given a base model pair $U_{base}$ and $X_{base}$, an initial draped shape $X_{init}$ is computed on the estimated body $B$, with reference to $U_{base}$. The open contours on both the simulated and the target meshes are extracted, associated with their respective counterparts, and the distances between them are measured along the skeleton of the underlying body. These longitudinal distances, together with the difference in circumference, are used to guide the relocation of control points $\{c^{i}\}$ on the open curves and others in the pattern. An algorithmic description is given in Algorithm 2 of Supplementary material.

\begin{figure}[htp]
\centering
\includegraphics[width=1.0\linewidth]{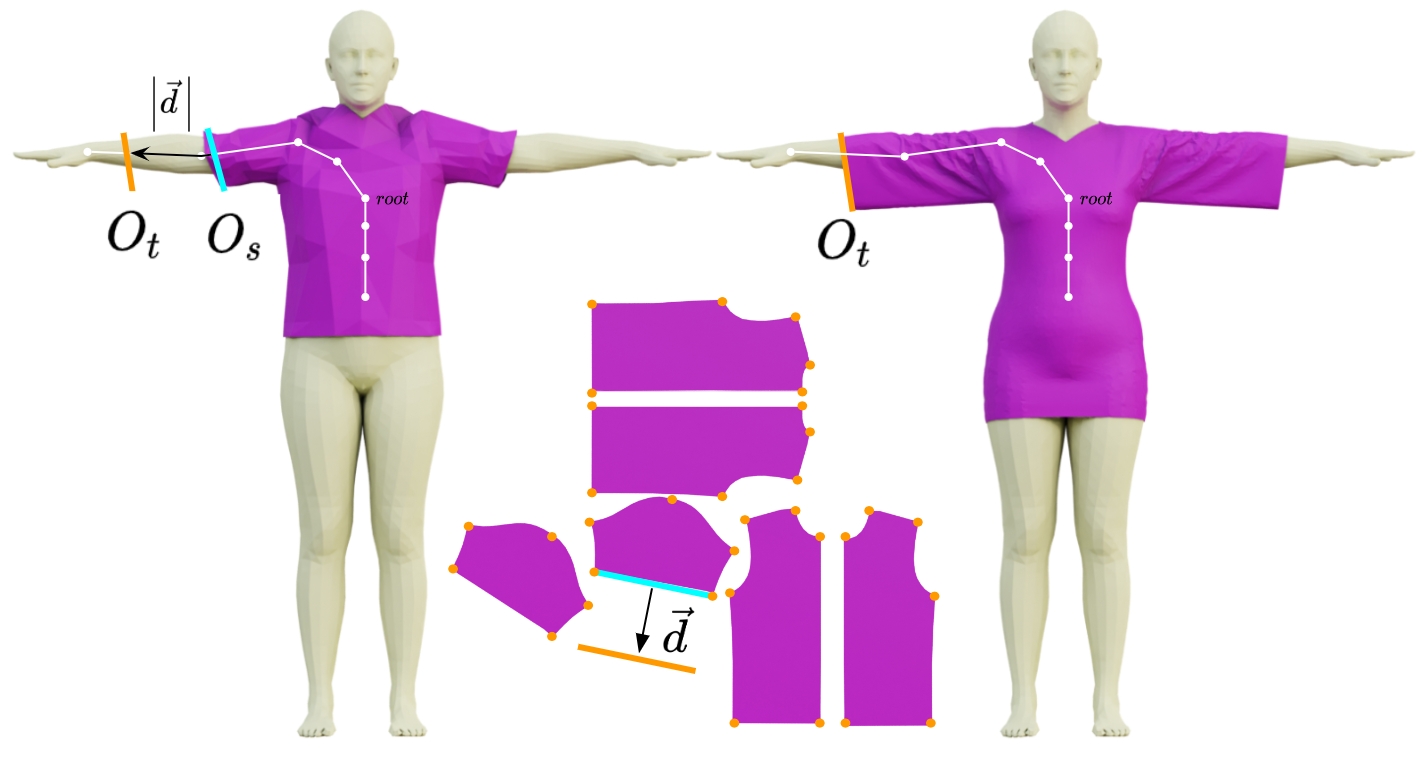}
\caption{An example of linear grading. The axial distance $\overrightarrow{\vec{d}}$ between open contours from the source mesh (left) to the target mesh (right) is used to displace the corresponding open curve (in blue) in the 2D panel.}
\label{fig:lg}
\end{figure}

In \Cref{fig:lg}, we illustrate an example of how the measurements on the cuff in 3D are used in the editing of 2D panel curves. The distance between the target cuff $O_t$ and the source $O_s$ measured along the arm bones, together with the difference in their circumferences, determine the amount of the displacement $|\vec{d}|$ of control points on the corresponding open curve (in blue) on the panel. The direction of $\vec{d}$ is derived by computing and normalizing the midpoint of the two endpoints of the open curve minus the average of other control points, while the width-changing vectors are computed by subtracting each control point from the other and normalizing the resulting vectors individually.

The above process could potentially lead to substantial location changes of control points, leading to undesirable topological distortion such as fold-over. To preserve the initial topology of the pattern mesh as well as the neighboring relationship among vertices,  we employ the 2D deformation method based on Mean Value Coordinates (MVC), similar to \cite{meng2012computer}. After the control points change their positions, the positions of other boundary vertices are updated in a per-curve manner, by repositioning each of them in the same relative location in a local coordinates system whose principal axis is defined by a vector connecting two end (control) points of the curve. The remaining interior vertices are then updated iteratively with reference to the deformed boundary position constraints by MVC. The edit of 2D panels is followed by a remeshing and draping process to have a simulated 3D garment, to reflect the change also in 3D. The resulting pattern $U_{LG}$ and its corresponding draped garment mesh $X_{LG}$ serve as a good initial state for the subsequent optimization-driven pattern alteration, which is described in \Cref{sec:opt}.

\subsection{Inverse garment simulation}
\label{sec:sim}
The garment-pattern result obtained from the previous phase is only an approximation of the target geometry. In the next phase, we further refine the pattern $U_{LG}$ through an optimization tightly coupled with a differentiable cloth simulation. Specifically, we extend the differentiable ARCSim\cite{liang19,narain2012adaptive} by revisiting both the dynamic solve and body-cloth interaction.

\subsubsection{Differentiable cloth simulation}
\label{diff}
At each forward simulation step, the draping garment over the estimated body is computed, taking into account external and internal forces until an equilibrium is achieved. The implicit Euler integration involves solving a linear system for the cloth motion, which is written as:
\begin{equation}
\label{eqn:linear}
(M-\Delta t^2J)\Delta v = \Delta t( f +  v J \Delta t),
\end{equation}
where $f$ is the sum of external forces (gravity, contact force) and internal forces (stretching, bending, etc). $M$ is the block diagonal mass matrix composed of the lumped mass of each node, and $J=\frac{\partial f}{\partial x}$ is the Jacobian of the forces. At each time step $ \Delta t$, we solve \cref{eqn:linear} for $\Delta v$ and update the velocity $v$ and position $x$. The equation could be written as $\hat{\mathbf{M}}\mathbf{a}=\hat{\mathbf{f}}$ for simplicity.

After the forward simulation with a predefined number of time step (10 to 20 in our experiments), a loss $\mathcal{L}$ (\Cref{sec:opt}) is measured between the simulated garment geometry and the target cloth mesh segmented from the 3D input. The error is used to back-propagate gradients to optimize the garment rest shape in terms of pattern parameters. Taking the implicit differentiation from \cite{liang19}, we use the analytical derivatives of the linear solver to compute $\frac{\partial \mathcal{L}}{\partial \hat{\mathbf{M}}}$ and $\frac{\partial \mathcal{L}}{\partial \hat{\mathbf{f}}}$ with the gradients $\frac{\partial \mathcal{L}}{\partial \mathbf{a}}$ backpropagated from $\mathcal{L}$. 

\subsubsection{Material model}
\label{mat}
We employ the linear orthotropic stretching model \cite{sifakis2012fem} to quantify the extent of planar internal forces in response to cloth deformation. The model defines the relation between stress $\sigma$ and strain $\epsilon$ 
using a constant stiffness matrix $\boldsymbol{H}$: $\sigma = \boldsymbol{H}\epsilon$, where
\begin{equation}
\begin{aligned}
\boldsymbol{H}&=\left[\begin{array}{lll}
H_{00} & H_{01} & 0\\
H_{01} & H_{11} & 0\\
0& 0 & H_{22}
\end{array}\right].
\end{aligned}
\end{equation}

\noindent The bending forces are modeled with piecewise dihedral angles which describe how much the out-of-plane forces would be when subject to cloth bending, as used in \cite{bridson2005simulation}:
\begin{equation}
\mathbf{f}_{i}=k \frac{\|\mathbf{e}\|^{2}}{\left\|\mathbf{A}_{1}\right\|+\left\|\mathbf{A}_{2}\right\|} \sin \left(\frac{\pi-\alpha}{2}\right) \mathbf{u}_{i}, 
\end{equation}
where $\alpha$ is the dihedral angle, $\mathbf{e}$ is the edge vector, $\mathbf{f}_i$ represents the bending force applied on the $i$-th vertex ($i$=1,...,4), $\mathbf{A}_1$ and $\mathbf{A}_2$ denote the areas of two triangles, $\mathbf{u}_i$
is the direction vector of the $i$-th node, and $k$ is the bending stiffness coefficient. 

\subsubsection{Acceleration of force vector/Jacobian matrix assembly}
\label{gpuacc}
ARCSim \cite{narain2012adaptive,liang19} uses the traditional approach of directly solving the linear system after the assembly of the extended mass matrix $\hat{\mathbf{M}}$ and the force vector $\hat{\mathbf{f}}$. 
The internal forces exerted by a triangle element to its nodes are split and accumulated to the global force vector, 
where the contributions from multiple adjacent elements are summed up for each node. Such force vector assembly process 
incurs a considerable overhead cost as the number of time steps grows. It is even more expensive for the extended mass matrix assembly as it contains the Jacobian of forces, 
which is large and sparse.  
We propose an efficient method for accelerating the assembly. As the same assembly is executed for each time integration, we exploit the fact that the inherent topological structure remains unchanged during the simulation, with a sequence of triangle elements in a fixed filling order. We encode this information in the form of a static mapping matrix, which converts the assembly process to a matrix multiplication, 
which is parallelizable on a GPU. Additionally, the matrix remains very sparse regardless of the mesh resolution, for which multiple numerical tools are available. 

\begin{figure*}[ht]
\centering
\includegraphics[width=0.95\linewidth]{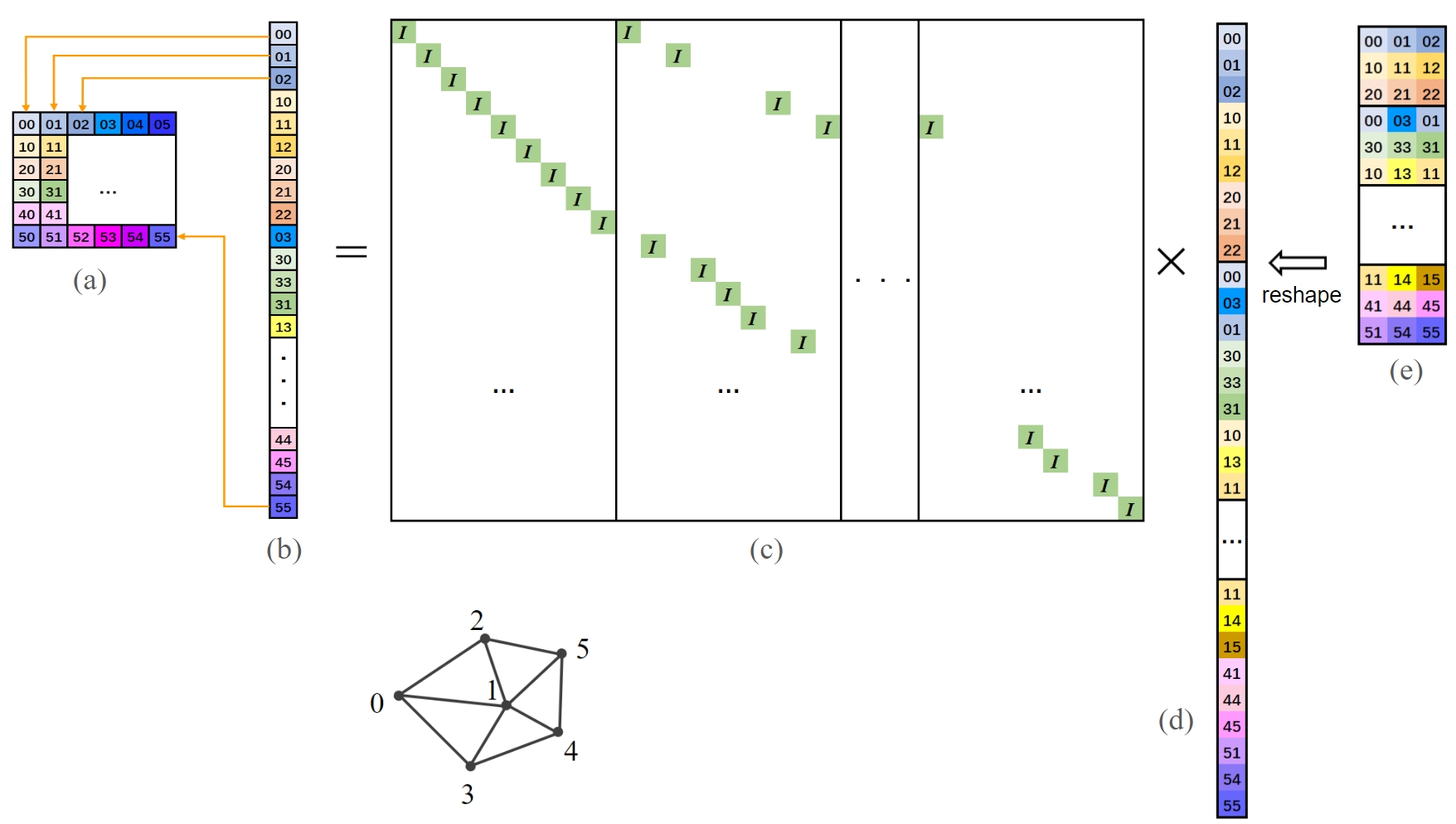}
\caption{
The assembly of a compressed Jacobian vector (b) is obtained  by a
static matrix multiplication,  
encoding the mapping from the per-element Jacobian (e) to the per force-node Jacobian (b). Compressed Jacobians are then transferred to the extended mass matrix (a) .}
\label{fig:jacobian}
\end{figure*}

In particular, the matrix $\hat{\mathbf{M}}$ is constructed by aggregating the local contributions from individual elements into the corresponding locations. Each triangle element (element hereafter) yields a Jacobian matrix of nine partial derivatives of the force with respect to the position of a node ($\frac{\partial f_{i}}{\partial x_{j}} $), $i,j$=1,2,3.
The Jacobians for all elements are packed into a \textit{Jacobian stack} as illustrated in \Cref{fig:jacobian} (e), where we use $mn$ to denote $\frac{\partial f_{m}}{\partial x_{n}}$ ($m$, $n$: global indices) for a compact representation. 

In ARCSim \cite{liang19, narain2012adaptive}, these Jacobians are assembled to $\hat{\mathbf{M}}$ through a total of $F\times 3\times 3$ assignment or addition operations, where $F$ represents the number of triangles in the garment mesh. Optimizing this process becomes crucial, especially considering its higher computational cost compared to the force vector assembly. To this end, we propose to realize the Jacobian assembly using matrix multiplication. $\hat{\mathbf{M}}$ is a sparse matrix, and moreover, directly representing the mapping from the Jacobian stack to the matrix form is not feasible, although it would be ideal for leveraging GPU-accelerated matrix multiplication. We address this issue by introducing an intermediate data structure called \textit{compressed Jacobian vector} (\Cref{fig:jacobian}(b)). It is a set of Jacobians for each force-node combination, which is obtained by first reshaping the Jacobian stack into a Jacobian vector (\Cref{fig:jacobian}(d)), and by encoding the mapping from the per-element Jacobian to the compressed Jacobian vector as a static mapping matrix (\Cref{fig:jacobian}(c)). Then a GPU-based sparse matrix multiplication is performed, effectively substituting the iteration-based Jacobian assembly. It is highly backpropagation-friendly, resulting in a considerable acceleration of the assembly process of linear solve(\Cref{tab:mapping}). Finally, the Jacobians in the compressed vector are transferred to $\hat{\mathbf{M}}$ (\Cref{fig:jacobian}(a)). The number of operations reduces to $N+2 \times E$ ($N$: number of nodes, $E$: number of edges), compared to the original $F \times 3 \times 3$. The assembly of batched elementary forces to the global force vector follows the same principle. 
 
\subsubsection{Efficient body cloth interaction}
\label{sdf}
One important component for draping simulation lies in the body-garment interaction, which involves the contact force computation and the garment-body collision handling, for which many algorithms have been proposed \cite{bridson2005simulation,tang2010, harmon2008}. In particular, the one based on non-rigid impact zones \cite{harmon2008} has been made differentiable by Liang et al \cite{liang19}. However, it remains computationally expensive, leading to rapid growth of the computation graph (i.e. memory-hungry) during forward simulation, and struggles to accommodate high-resolution meshes. Hence, we chose to compromise by implementing a light collision handling scheme that makes use of the signed distance function (SDF).

When the signed distance of a query garment vertex $x$ falls below a threshold (indicating proximity to the body), the repulsion force is triggered between the body and the garment, with its magnitude inversely proportional to their distance. We use the classical Coulomb’s model for friction force, which is elicited when there is relative movement along the surface tangent. While the repulsion forces prevent the interpenetration, occasional collisions might still occur and need correction after the dynamic simulation. To this end, for any garment vertex $x$ with $sdf(x)<0$ we present the collision resolving setup, correcting the interpenetration by:
\begin{equation}
\tilde{x}= x+(\delta -sdf(x))\cdot \mathbf{n}, 
\end{equation}
where $\mathbf{n}=\frac{\nabla sdf(x)}{\left|\nabla sdf{x}\right|}$ is the spatial gradient of $sdf(\cdot)$ (also the surface normal), and $\delta$ denotes the collision thickness. This scheme significantly enhances the speed of both forward and reverse simulation while maintaining the performance level.

The signed distance of the underlying body is computed either by a trained neural network, which is naturally differentiable, or analytically on the fly with Kaolin \cite{kaolin}). In the latter case, as it involves only unit operations, it can also be seamlessly integrated in the differentiable simulation. Since the SDF is also used to compute the repulsion forces for intersection prevention, the position correction (projection) is generally sparse and minor if there is any. As a result, it ensures smooth backpropagation without introducing discontinuities.

\begin{figure}[htp]
\centering
\includegraphics[width=1.0\linewidth]{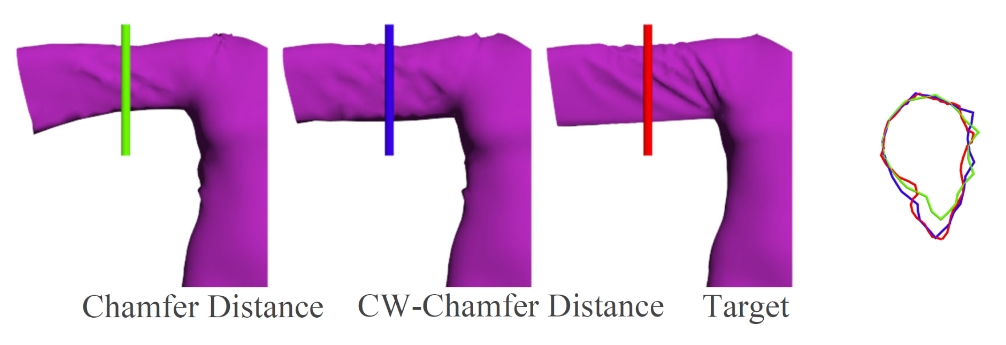}
\caption{The cross-sectional curves on the right, generated from three meshes on the left: two simulated meshes, and the target. 
The colors of the cutting planes are used to draw the cross-sections. The curvature-weighted Chamfer (blue) leads to a sleeve draping silhouette closer to the ground truth (red), compared to the standard Chamfer distance (green).}
\label{fig:Wchamfer}
\end{figure}

\subsection{Optimization-based pattern alteration}
\label{sec:opt}
In this phase, we further refine both the pattern state $U_{LG}$ and the simulated garment $X_{LG}$ obtained from the previous stage through optimization using the differentiable draping simulator. At each iteration, the simulated garment geometry $ X=\{x_i\}$ is compared with the target $T=\{t_i\}$ using a loss function, subsequently utilized by a gradient-based algorithm to refine the pattern shape. We define the following loss over the effective control points $K$ and the physical parameters $\Gamma$:
\begin{equation}
\begin{aligned}
\mathcal{L} &=\mathcal{L}_{rec}(X=\textit{Sim}(U(K),\Gamma; \textit{SMPL}(\theta,\beta)),T)\\&+\lambda_{seam} \mathcal{L}_{seam}(U(K)),
\end{aligned}
\end{equation}
where $\lambda$'s are weights. It combines the reconstruction loss $\mathcal{L}_{rec}$ and the seam-consistency loss $\mathcal{L}_{seam}$ penalizing the inconsistent curve lengths along the seam. The reconstruction error is composed of the Chamfer distances $\mathcal{L}_{CF}$ measured both between the surfaces ($X$ and $T$) and among open contours ($X_{open}$ and $T_{open}$), 
\begin{equation}
\mathcal{L}_{rec}=\mathcal{L}_{CF}(X, T)+\lambda_{open} \mathcal{L}_{CF}(X_{open},T_{open}) + \lambda_{mat} \mathcal{L}_{mat}(\Gamma).  
\end{equation}

The Chamfer distance, widely used for fitting deformable surfaces, has proven to work well in most cases. However, we observed that it is not sufficient for certain garment targets, due to its “myopia” that each point only considers its nearest neighbor on the other mesh, neglecting the surroundings. In regions with high curvatures, often present in the folded geometry of loose clothes, this can lead to lower geometric accuracy (See \Cref{fig:Wchamfer}). Hence, we use curvature-weighted Chamfer distance \cite{bongratz2022vox2cortex} instead, which prioritizes high-curvature regions, subsequently improving the reconstruction of densely folded regions: 
\begin{equation}
\begin{aligned}
\mathcal{L}_{CWCF}\left(X, T\right) &=\frac{1}{\left|X\right|} \sum_{x \in X} \kappa(\tilde{t}) \min _{t \in T}\|x-t\|^{2}
\\&+\frac{1}{\left|T\right|} \sum_{t \in T} \kappa(t) \min _{x \in X}\|t-x\|^{2}, 
\end{aligned}
\end{equation}
with $\kappa$ the mean curvature and $\tilde{t}=\arg \min_{y \in T} {(\|x-y\|)} $. The seam loss serves as the regularization that guarantees the consistent curve lengths of two panels along the seam:

\begin{equation}
L_{seam}(U)=\sum_{i}^{|S|} \sum_{j}^{|E_{i}|=|{E_{i}^{corr}}|} \left (|e_i^{j}|-|e_i^{corr, j}|\right )^{2} , 
\end{equation}
\noindent where $S=\{S_i\} \subset X$ denotes the set of 3D seam curves, and $E_i, E_i^{corr} \subset U$ are the sets of edges $e^j=u^j-u^{j+1}$ along the 2D panel curves comprising the seam counterparts of $S_i$.

We also optimize over the physical parameters $\boldsymbol{H}$ and $k$ by adding them into the variable set:
\begin{equation}
\Gamma :=(H_{00},H_{01},H_{11},H_{22},k).
\end{equation}
To penalize unrealistic material parameter combinations, we constrained the elements within the physically plausible ranges(above a non-negative threshold 1e-6), with $\mathcal{L}_{mat}=\text{relu}(1e^{-6}-\Gamma)$. These physical parameters are added to the variable set in later iterations, once the pattern shapes reach an approximate optimum.

\section{Experiments}
\label{sec:exper}

\subsection{Implementation details}
We now describe the main implementation details. Further information is provided in the supplementary material. 

\noindent \textbf{Simulation.}
We set one time step $\Delta t$ to 0.05s, and the number of time steps for one forward simulation between 10 and 20. The garment resolution tested ranges from 1K to 8K vertices. We set the collision thickness to the conventional value 1e-3m. A variant of DeepSDF \cite{Mu2021ASDF} with periodic activation functions \cite{sitzmann2020implicit} that learns the SDF of the wearer's body has been used for collision handling. By vectorizing as much as possible the force and jacobian computation, our extension to ARCSim differentiable cloth simulator \cite{liang19} allows it to run all computations on a GPU. \Cref{tab:mapping} summarizes the computation time of linear solve by the baseline model \cite{liang19} and ours, measured on a NVIDIA GeForce 3090, for T-shirt garments with different resolutions (1K to 3K).
Note that in our model the first iteration involves the construction of the deterministic mapping, which is reused in subsequent iterations. In contrast, the baseline model requires the overhead of assembly for every iteration. The acceleration of the linear solver and the SDF-based collision handling scheme result in 
a 14 times overall speedup compared to \cite{liang19}. 

\newcolumntype{d}[1]{D{.}{.}{#1}}
\begin{table}[ht!]
\centering
\caption{The computational time for the matrix assembly and linear solve (in seconds) using the baseline model and ours, measured on a T-shirt garment with varying resolutions. Note how our method improves the speed as the number of iterations increases, especially during the reverse process.}
\scalebox{0.8}{
\begin{tabular}{l d{3.3} d{3.3} d{3.3} d{3.3} l}
\noalign{\smallskip}
      & \multicolumn{2}{c}{Baseline\cite{liang19}} & \multicolumn{2}{c}{Ours}  \\
  \cmidrule(r){2-3}  \cmidrule(r){4-5}
\backslashbox{\# verts}{iterations}  & 1  & 20  & 1   & 20  & speedup\\
\hline
1K (forward)&  2.8 &  83.7&  19.1 & 59.4  & 1.4x\\
1K (reverse) & 7.0 &  155.6 &  2.7 & 60.0  & 2.59x\\
\hline
2K (forward)& 8.4  &  178.1 &  67.6 & 133.0 & 1.34x\\
2K (reverse)&  17.2 &  466.0 & 6.3  & 128.7 & 3.62x\\
\hline
3K (forward)& 12.3 & 257.9  & 133.6  & 228.6 &1.13x \\
3K (reverse)& 24.7 & 505.4  &  12.0 & 186.3 &2.71x\\
\noalign{\smallskip}
\hline
\end{tabular}
}
\label{tab:mapping}
\end{table}

\noindent \textbf{Optimization.} 
The variables are optimized by using the Adam optimizer \cite{kingma2015adam}, with a learning rate $10^{-3}$, $\beta_1$ = 0.9, $\beta_2$ = 0.999. We empirically set $\lambda_{open}=0.1$, $\lambda_{mat}=0.01$ and $\lambda_{seam}=0.01$. We observed that the addition of material parameters noticeably impacts the result. Conversely, excluding these parameters from optimization might lead to distorted panels. We present the results of a related ablation study in the following section. The material parameters are initialized with a material set selected by the user from the material library, which spans a range of common fabrics. 
Additionally, the pattern variables are effectively initialized through linear grading. 


\subsection{Quantitative and qualitative comparisons}

\begin{figure}[htp]
\centering
\includegraphics[width=1.0\linewidth]{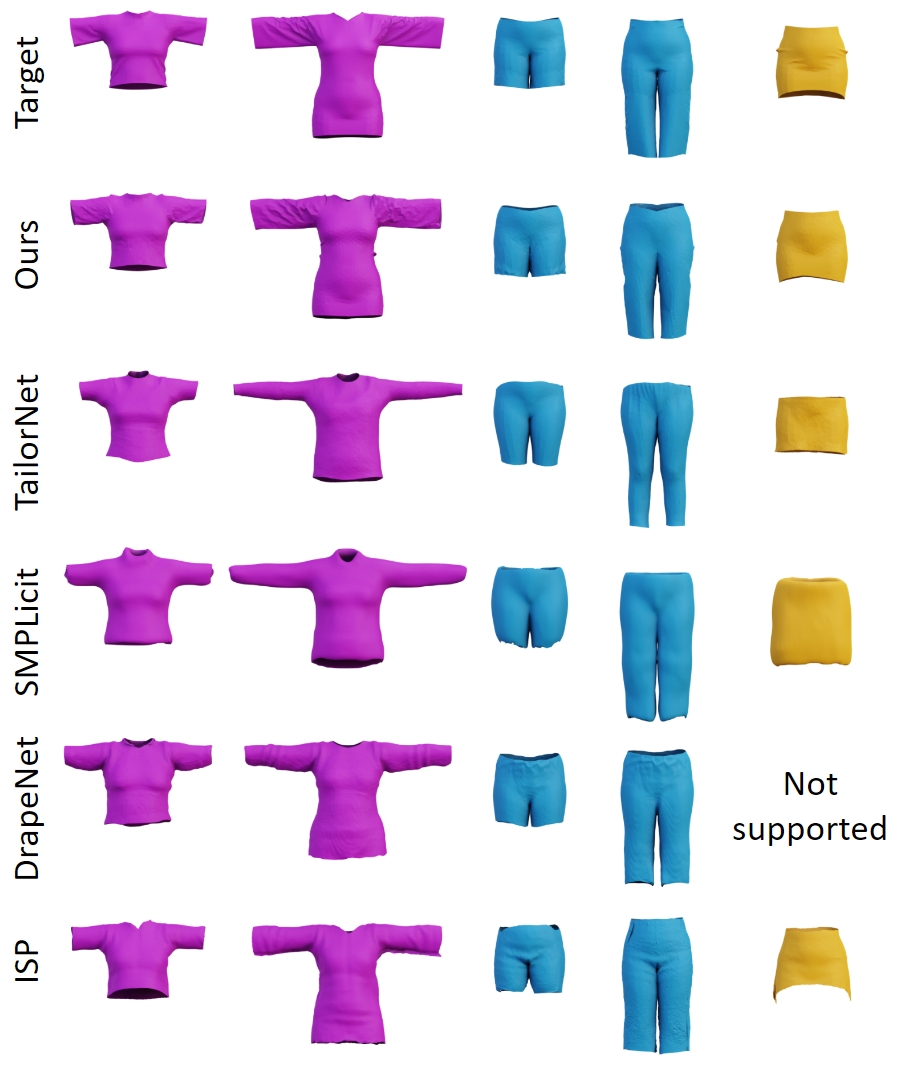}
\caption{
Comparison of 3D garment reconstruction of our method with others \cite{patel20tailornet, corona2021smplicit,de2023drapenet,li2024isp}. Our method reproduces faithful garment shapes, even accounting for intricate geometry details like wrinkles on large sleeves. Best viewed on screen zoomed-in.
} 
\label{fig:3deval}
\end{figure}

We evaluate our model on a number of representative garment types and compare it with previous works. We first evaluate the performance on 3D garment reconstruction, and then on 2D pattern estimation. To carry out a fair comparison, we use the garment meshes of a third-party dataset \cite{KorostelevaGarmentData} as targets (i.e. test data), which is an unseen dataset for all evaluated methods. It is also one of the rare datasets that provide 2D sewing patterns for every 3D garment mesh, enabling the evaluation of our results in both 3D reconstruction and 2D pattern estimation. We describe the detailed results below.

\noindent \textbf{3D garment reconstruction.} 
We compare our approach to the related methods for garment fitting, and utilize 3D garment geometry as targets. We run the Adam optimizer \cite{kingma2015adam} for a varying number of iterations until the convergence for each method, while noting that the parameter space representing the garment geometry differs among them: The coordinates of effective control points and material parameters for our method, the $\gamma$ garment style parameter for TailorNet \cite{patel20tailornet}, the latent vector $z=[z_{cut},z_{style}]$ describing garment cut and style for SMPLicit \cite{corona2021smplicit}, and the latent codes $z$'s encoding the garment characteristics in DrapeNet and ISP\cite{de2023drapenet,li2024isp}. The results are quantitatively evaluated using two metrics: Chamfer distance to the ground truth mesh vertices, and the angular error to measure the similarity of the computed normal vectors, similar to \cite{bednarik2020shape}. As shown in  \Cref{tab:3drecon}, our method is consistently better than others, which is confirmed by qualitative results, shown in \Cref{fig:3deval}. We observe that the performance of the data-driven approaches is biased by the training dataset. It is clear that TailorNet basically has learned over tight-fit datasets, so it does not generalize very well when fitting to loose styles, as seen in the Pants example. In contrast, our approach reconstructs accurate 3D geometry, for both loose and tight garments. Drapenet uses unsigned distance functions (UDFs) and requires extra computations for meshing, and it is sensitive to the initialization of latent code for the optimization.

\begin{table*}[ht!]
\centering
\caption{Quantitative evaluation in 3D (garment reconstruction) and 2D (pattern estimation). 
To measure the accuracy of 2D patterns, we evaluate the turning function metric for comparing polygonal shapes \cite{turningfunction} and the surface error (the average of normalized surface difference error computed for each patch).
}
\renewcommand{\arraystretch}{1.1}
\begin{tabular}{lcccccccc} 
\noalign{\smallskip}
      & \multicolumn{5}{c}{3D Reconstruction} & \multicolumn{3}{c}{2D Estimation} \\
\noalign{\smallskip}
      & \multicolumn{5}{c}{Chamfer distance / Normal similarity} & \multicolumn{3}{c}{Turning /Surface area} \\
\cmidrule(r){2-6} \cmidrule(r){7-9}
Garments & SMPLicit  & TailorNet & Drapenet & ISP & Ours & NeuralTailor & SewFormer & Ours   \\
\hline
T-shirt &1.4 /- & 0.331 / 0.081 & 0.689 / 0.129 & 0.297 / 0.094
&\textbf{0.112} / \textbf{0.049}&10.50 / 0.13 & 10.46 / 0.15
&\textbf{9.12} / \textbf{0.09}
\\
Dress&3.2 / -& 1.305 / 0.161 & 0.619 / 0.135 & 0.189 / 0.131
& \textbf{0.110} / \textbf{0.075}&11.20 / 0.37 & 11.77 / 0.43
&\textbf{10.96} / \textbf{0.10}
\\
Shorts&1.3 / -&  1.036 / 0.050 & 0.131 / 0.048 & 0.202 / 0.095
& \textbf{0.126} / \textbf{0.043} &\ 7.41 / 0.05 & 7.74 / 0.06
&  \textbf{7.33} / \textbf{0.04}
\\
Pants&2.9 / -& 2.587 / 0.104 & 0.485 / 0.085 & 0.185 / 0.077
&\textbf{0.142} / \textbf{0.049}&\textbf{6.77} / \textbf{0.01} & 6.79 / 0.20 
&  7.00 / 0.08
\\
Skirt&6.5 / -&1.300 / 0.063 & - / - & 0.435 / 0.093
&\textbf{0.106} / \textbf{0.014}& 4.31 / 1.11 & 4.42 / 0.62 
&\textbf{4.14} / \textbf{0.04}
\\
\noalign{\smallskip}
\hline
\end{tabular}
\label{tab:3drecon}
\end{table*}

\begin{figure}[htp]
\centering
\includegraphics[width=1.0\linewidth]{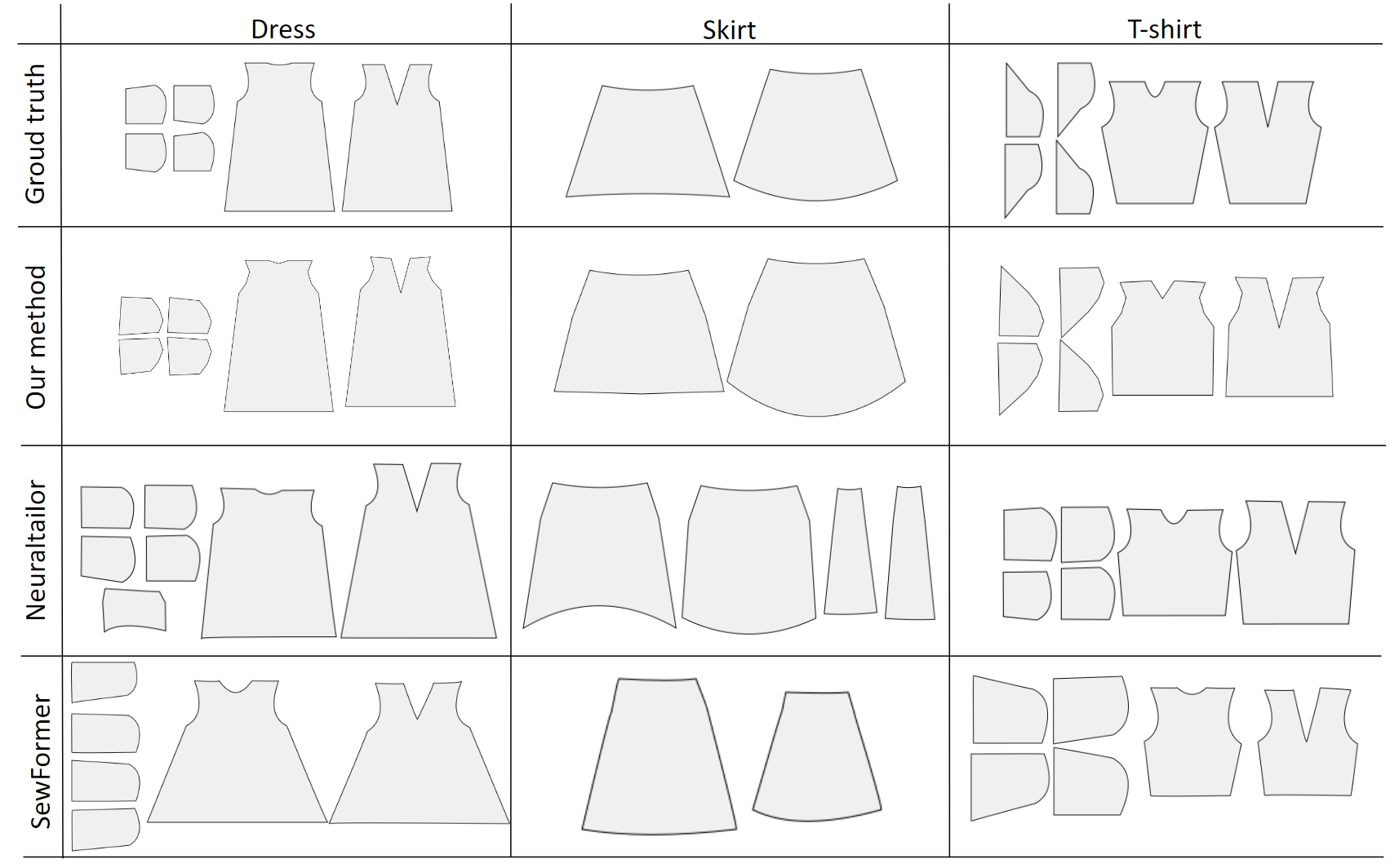}
\caption{2D pattern estimation from 3D garment meshes. From top to bottom: Ground truth, ours, NeuralTailor\cite{korosteleva2022neuraltailor} and SewFormer\cite{liu23sewformer}. 
}
\label{fig:2deval}
\end{figure}

\noindent \textbf{2D Pattern Estimation}. The quantity of research focusing on sewing pattern recovery directly from a given 3D input data is rather limited, with the majority of them dedicated to precise but minor adjustments to existing patterns \cite{wang18,bartle16}. We compare our work with NeuralTailor \cite{korosteleva2022neuraltailor} and SewFormer\cite{liu23sewformer}, two deep learning frameworks that predict a structural representation of a sewing pattern from a 3D point cloud and a 2D image, respectively. To facilitate comparison  with the ground-truth patterns, the experiments were conducted under favorable conditions for their work -- We selected five patterns from the dataset used in NeuralTailor as the ground-truth ones. Then, we generate 3D drape shapes at a T-pose by using an independent simulator \cite{3dsMax}, differing from both ours and theirs. To evaluate SewFormer, we render 2D images with settings similar to those used by SewFormer. Some of the results are illustrated in \Cref{fig:2deval}. We observe that their method makes very good predictions on the trouser-like garments as the geometric variation of pants and shorts are limited and well covered in their training dataset. For the other garment types, however, our method produces better results. To quantitatively measure the quality of estimated 2D patterns, we have used two metrics: (1) the turning function metric for comparing polygonal shapes \cite{turningfunction}, and (2) the relative error in surface area,  as determined by averaging normalized surface difference error $\frac{1}{\left | P \right | }\sum \frac{\Delta A(P_i)}{A(P_i)}$ computed for each panel $P_i$.

\begin{figure}[htp]
\centering
\includegraphics[width=1.0\linewidth]{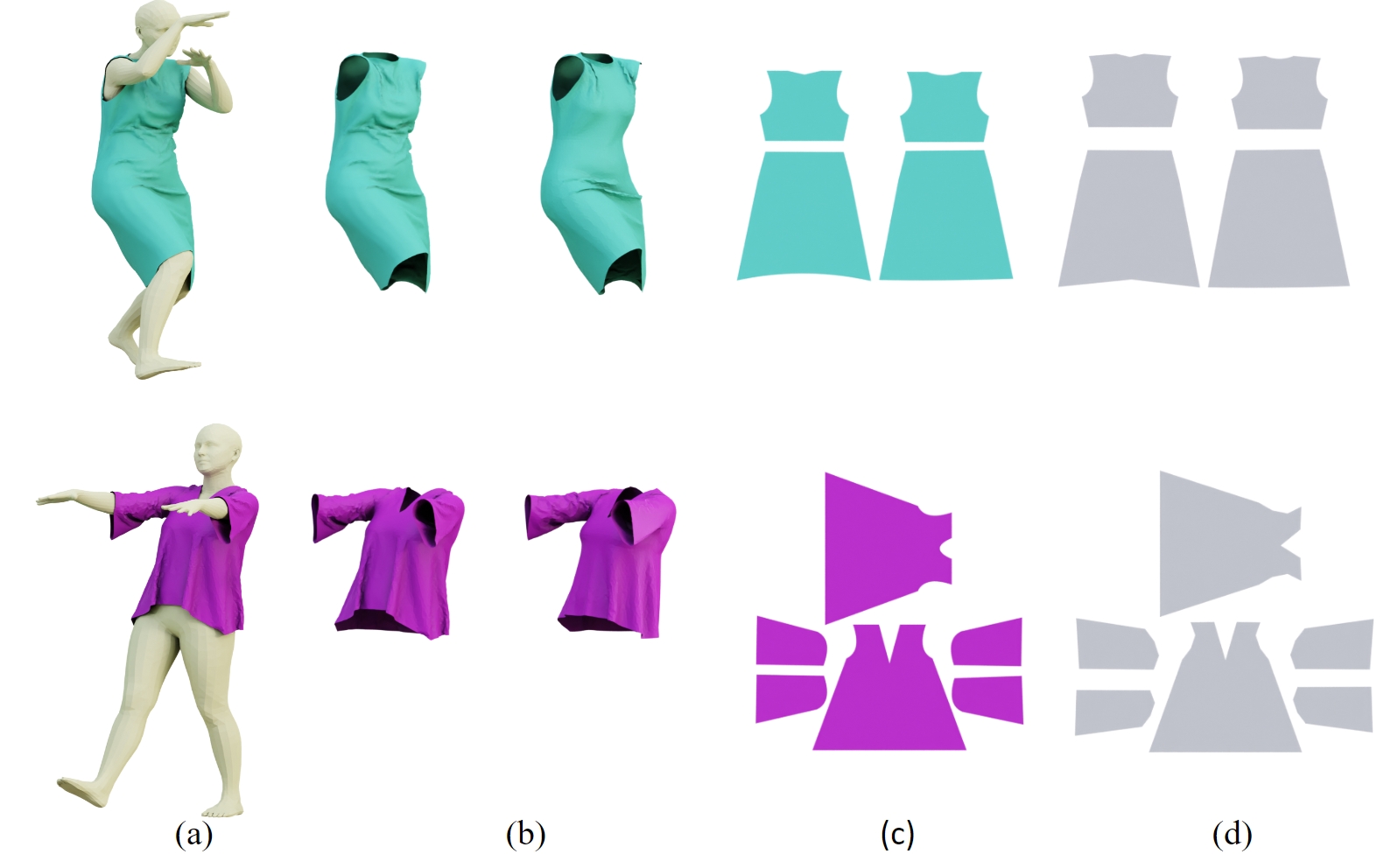}
\caption{Results of our method evaluated using varying poses. (a) Target clothed body; (b) target garment mesh (left) and the reconstructed mesh (right); (c) ground-truth pattern; and (d) our estimated pattern.}
\label{fig:variations}
\end{figure}

\noindent\textbf{Generalization to different poses and shapes.}
Target garment shapes draped on a T-posed body allow for a fair comparison to NeuralTailor, since it has been trained on garments in this setting. To demonstrate that our method also performs well in other settings, we have tested our method on some example meshes from SewFormer \cite{liu23sewformer}. As shown in \Cref{fig:variations}, our method is able to faithfully reconstruct 3D garments worn by the individuals in challenging poses, while producing consistent patterns close to the ground truth.

\subsection{Recovery of physical parameters}
\begin{figure}[h!]
\centering
\includegraphics[width=0.9\linewidth]{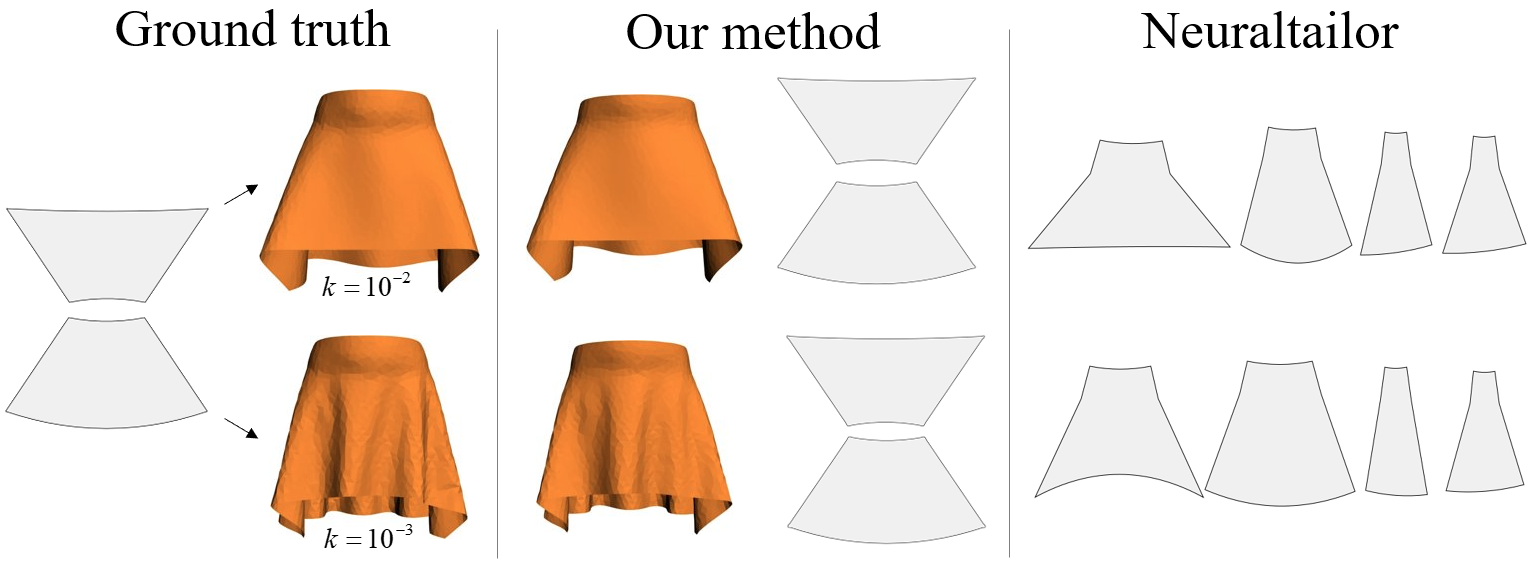}
\caption{Sewing patterns estimated from two input meshes, both simulated from an identical ground-truth pattern but with varying bending coefficients. By optimization over bending coefficient, our method correctly finds the panel shapes, compared to the alternative method.}
\label{fig:bend}
\end{figure}

To demonstrate the capability of our method to faithfully recover physics, two draping skirt meshes were simulated using identical sewing patterns but varying only the physical parameters. Then we used them as targets and compared our estimated patterns with those generated from NeuralTailor \cite{korosteleva2022neuraltailor}. As shown in \Cref{fig:bend}, our method can faithfully capture 3D garment geometric variations originating from different bending parameters, while producing consistent patterns close to the ground truth. On the contrary, NeuralTailor translates the geometric variation into that of panels, yielding a significantly different pattern for each target instance. 

\subsection{Evaluation on 3D scan data and retargeting}
\begin{figure}[htp]
\centering
\includegraphics[width=1.0\linewidth]{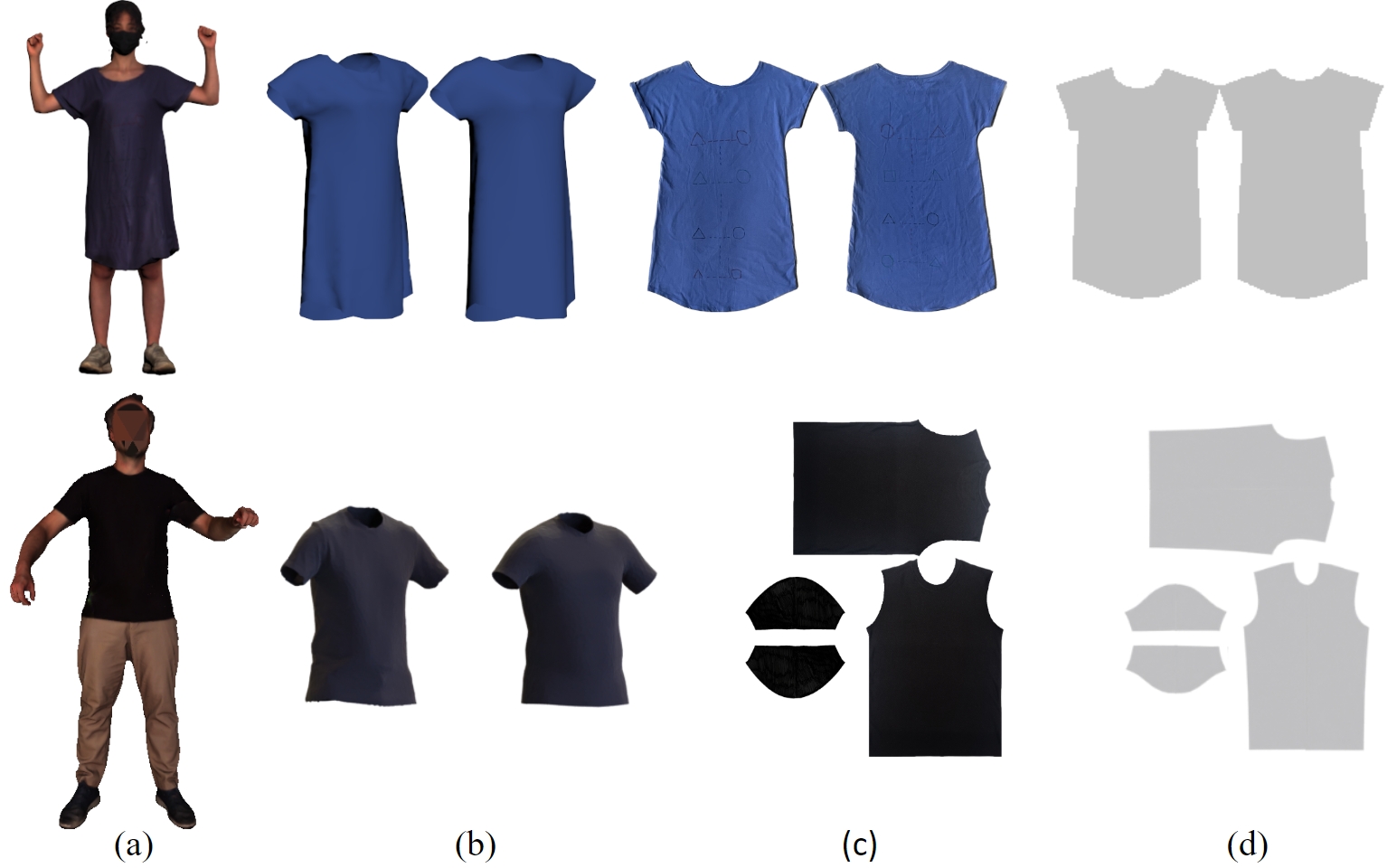}
\caption{Results of our method evaluated using 3D scan data. (a) Input 3D scan; (b) segmented target (left) and simulated garments (right); (c) ground-truth pattern; and (d) estimated pattern.
}
\label{fig:scan}
\end{figure}

\begin{figure}[htp]
\centering
\includegraphics[width=0.9\linewidth]{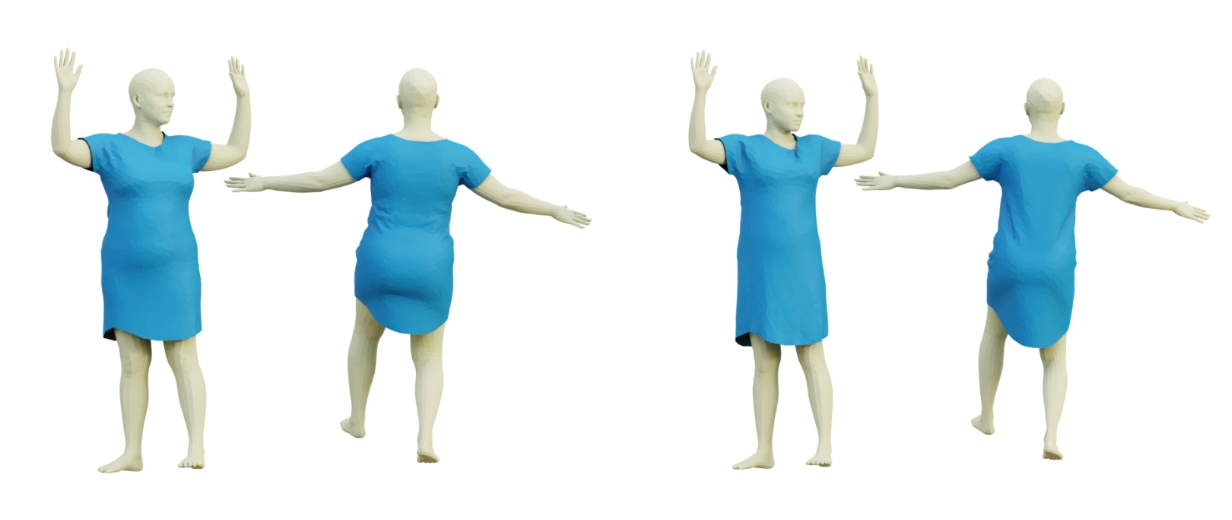}
\caption{The dress model obtained from the 3D scan (\Cref{fig:scan}) has been retargated to two new SMPL bodies.}
\label{fig:retarget}
\end{figure}

We conducted a qualitative evaluation of our method using 3D scans obtained with the Vitronic VITUS Human Solutions body scanner\cite{vitronic}, the captured meshes are of high quality and without holes. The ground-truth patterns have been obtained by placing transparent papers over the flattened garments, drawing along the seams, and then digitally cutting along the traced lines after scanning. As shown in \Cref{fig:scan}, our method outputs reasonable, quality estimations of the 3D garment and the 2D pattern. Since the recovered pattern is simulation compatible, it can be easily reused 
by a simulator to generate draping shapes on new conditions, as shown in \Cref{fig:retarget}. More results are provided in the accompanying video. 

\section{Ablation study}
\label{ablation}
Here we report the results of our ablation study, where we examine the contributions of individual components to the overall performance (\Cref{tab:ablation}). Our model achieved a Chamfer distance precision (CF) of 0.1103 (in \textit{mm}) and a cosine distance of normals (NOR) of 0.075. It outperforms other settings where the curvature-weighted Chamfer loss is replaced with the vanilla Chamfer (1st row of \Cref{tab:ablation}),  when the seam consistency loss term is removed (2nd row), or when the optimization of physical parameters was disabled (3rd row). These results confirm the importance of both loss terms and the integration of physical properties in the optimization process.

\begin{figure}[htp]
\centering
\includegraphics[width=1.0\linewidth]{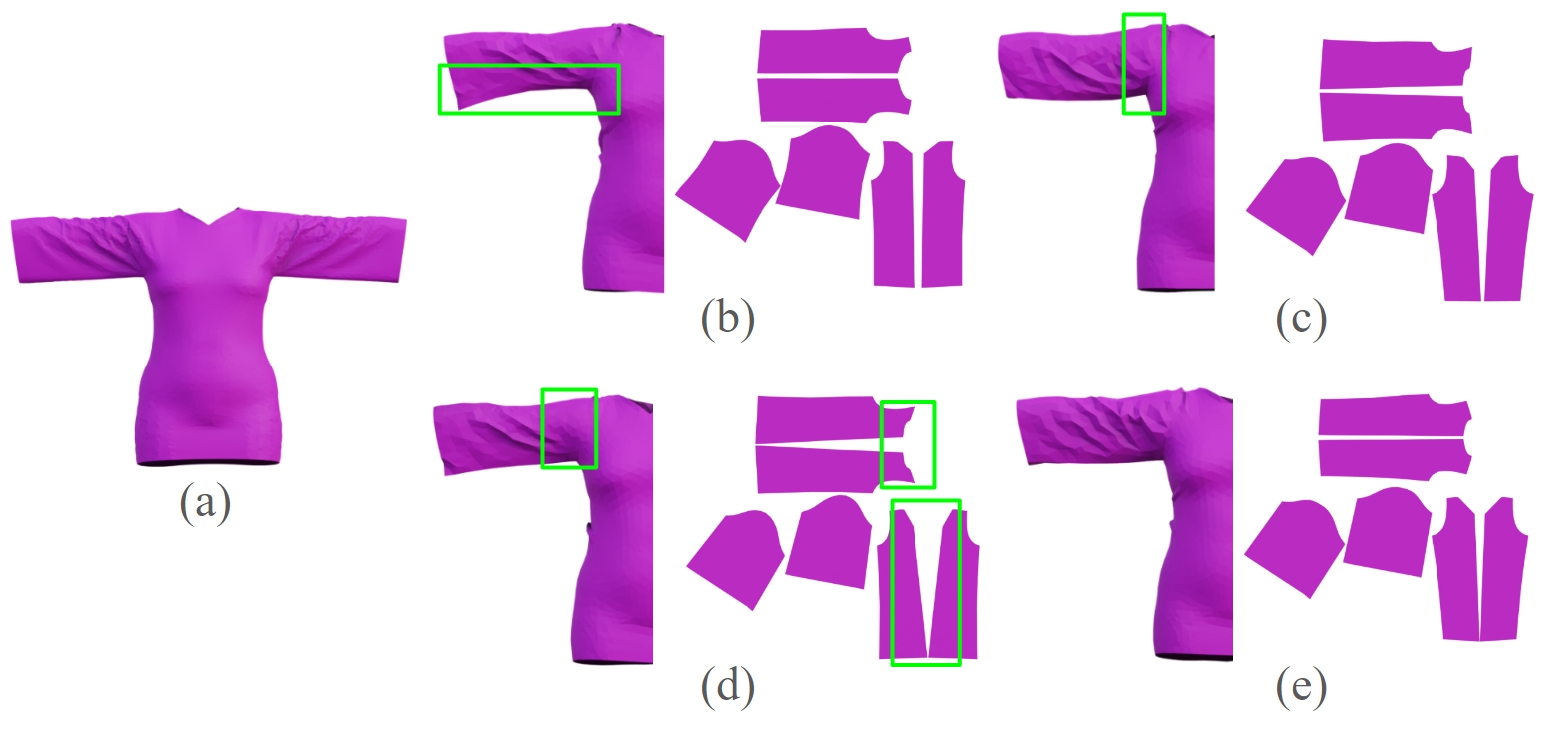}
\caption{Results of our ablation study on loss terms: target mesh (a); results with vanilla Chamfer loss (without curvature weights (b), without seam loss (c), and without physical parameters (d); our results (e).}
\label{fig:ablation}
\end{figure}

\Cref{fig:ablation} illustrates the reconstructed models obtained from the ablation study. We observe that our model (\Cref{fig:ablation}(e)) bears the closest visual resemblance to the target. The use of weighted Chamfer distance allows for better capture of the armpit region and the lower part of the sleeve (\Cref{fig:ablation}(b)). The absence of seam loss leads to a puckered seam around the shoulder, resulting from the extra tension exerted on the the shorter seam (\Cref{fig:ablation}(c)). The optimization of physical parameters helps to recover fine wrinkles, as well as more plausible pattern estimation. As we can see in \Cref{fig:ablation}(d), the ``pear shape'' of the body/dress has been solely attributed to the increasing panel width along the torso, when the material parameters were disregarded during the optimization.


\begin{table}[htb!]
\caption{Ablation study.}
\scalebox{0.9}{
\centering
\begin{tabular}{ lcc }
\hline
Method & Chamfer distance (CF)    & Normal similarity\\
\hline
Ours(w/o curvature CF) & 0.115    &  0.085\\
Ours(w/o seam loss) & 0.117 & 0.076  \\
Ours(w/o physics) & 0.113 & 0.081\\
Ours& \textbf{0.110} & \textbf{0.075}\\
\hline 
\end{tabular}
}
\label{tab:ablation}
\end{table}

\section{Conclusion}
\label{sec:con}
We have presented a method to recover simulation-ready garment models from a given 3D geometry of a dressed person. Basing our work on a differentiable simulator, we refine the 2D sewing pattern shape through inverse simulation, ensuring that the physically based draping of the corresponding sewn garment closely matches the given target. Our experimental results confirm that our system can produce simulation- and fabrication-ready patterns on a range of representative garment geometries, outperforming comparable state-of-the-art methods.

Our approach presents several limitations that suggest avenues for future exploration. First, the iterative optimization process involving forward and reverse simulation is time-consuming. 
Further acceleration can be employed to achieve faster convergence in optimization processes \cite{jang23wnn}. Second, although our linear grading scheme effectively adjusts the base model to align with the target prior to the optimization, the results can be sensitive to initial values, potentially resulting in different local minima. As well, predefined design choices such as mesh resolution and the identification of control points on the panels can also impact the outcome. 
Finally, our method is unable to handle cloth self-collision and multi-layer garment inputs, which we leave as future work.

\bibliographystyle{eg-alpha-doi} 
\bibliography{egbibsample}       


\newpage
\appendix 

\begin{algorithm*} 
\resizebox{0.9\textwidth}{!}{%
\begin{minipage}{\textwidth}
\textbf{Input:}\\
$C$ =
\{$C_p$\}, where $C_p$ is an ordered set of control points of panel $p$. \\
\textbf{Output:}\\
$K$: Effective control points \\
$InterSym$: A set of panel pairs in correspondence and the orthogonal matrix \{($C_p$, $C_{p'}$, $M_p$)\}, so that $c_{p}^i \in C_p$ corresponds to $c_{p’}^i\in C_{p'}$ and $(C_p \times  M) \simeq  C_{p'}$.\\
$IntraSym$: A set of control point pairs in mirror symmetry and the reflection matrix per panel \{$(Q,Q',M)$\} such that $(Q \times  M) \simeq  Q'$.\\
\renewcommand{\algorithmicrequire}{\textbf{Procedure:}}
\renewcommand{\algorithmicensure}{\textbf{return:}}
\caption{Pattern symmetry detection.}
\label{alg:sd}
\begin{algorithmic}
\Require SymmetryDetection($C$) 
\State $K$, $InterSym$ $\gets$ InterSymmetry($C$) 
\State $IntraSym$ $\gets$ $\emptyset$
\For {\textbf{each} $C_p \in K$}
\State $(Q, Q',M)$  $\gets$ IntraSymmetry($C_p$)

\If { $(Q, Q',M) \neq (Null)$}
\State 1. $IntraSym \gets IntraSym$ $\cup$ $(Q, Q',M)$ 
\State 2. $K \gets K \setminus Q'$
\EndIf
\EndFor
\Ensure $K, InterSym, IntraSym$
\renewcommand{\algorithmicrequire}{\textbf{Procedure:}}
\State
\Require InterSymmetry($C$)
      \State $K$ $\gets$ $C$, $InterSym$ $\gets$ $\emptyset$ 
      \For {$C_p$ \textbf{in} $K$}
       \State $error_p  \gets \infty $, $M_p \gets I$, $p' \gets -1$
      \For {$C_q$ \textbf{in} $K$}
      \State 1. $C_q^r$ $\gets$ ReversePointsOrder($C_q$) 
      \State 2. $error_{tmp}, M_{tmp}, C_{tmp} $ $\gets$ RigidAlignment($C_p, C_q^r$) 
      \If { $error_{tmp}<\epsilon$ and $error_{tmp}<error_p$} 
        \State $p'$ $\gets$ $q$, $error_p \gets error_{tmp}$, $M_p \gets M_{tmp}$,  $C_{p}$ $\gets$ $C_{tmp}$
      \EndIf
      \EndFor
      \If { $p' \neq -1$}
      \State 1. $InterSym \gets InterSym$ $\cup$ ($C_p$,$C_{p'}$,$M_p$) 
      \State 2. $K \gets K \setminus C_{p'}$
      \EndIf
      \EndFor
\Ensure  $K, InterSym$
\State
\Require IntraSymmetry($C_p$)
      \State $l$ $\gets$ $\left | C_p \right |$
      \For { $i= 0 ... l/2$}
      \State $U$ $\gets$ $ \left \{c_{p}^u\right\}$, $u=\left \{ i...i+\left \lfloor l/2 \right \rfloor -1  \right \} $
      \State $V$ $\gets$$ \left \{c_{p}^v\right\}$, $v=\left \{ (i+\left \lceil l/2 \right \rceil)  \% l...(i+l-1)\% l  \right\}$
      \State $V^{r}$ $\gets$ ReversePointsOrder($V$)
      \State $error$, $M$, $U'$ $\gets$ RigidAlignment($U, V^{r}$)
      \If { $error$ $<$ $\epsilon$}
      \State \textbf{return} ($U$,$U'$,$M$)
      \Else
      \State \textbf{return} ($Null$)
      \EndIf
      \EndFor

\State
\Require RigidAlignment($C_1, C_2$)
      \State $l$ $\gets$ $\left | C_2 \right |$,  $error_{min}$ $\gets$ $\infty$, $M^*$ $\gets I$ ,  $C_2^* \gets C_2$
      \For {$i = 0 ... l$}
      \State $C_2$ $\gets$ Concatenate($C_2[i:], C_2[0:i]$)
      \State $error$, $matrix$ $\gets$ OrthogonalProcrustes($C_1,C_2$)
      \If {$error < error_{min}$} 
      \State  $M^*$ $\gets$ $matrix$,  $ C_2^* \gets C_2$, $error_{min}$ $\gets$ $error$ 
      \EndIf         
      \EndFor
\Ensure  $error_{min}, M^*, C_2^*$
\end{algorithmic}
\end{minipage}%
}
\end{algorithm*}

\begin{algorithm*}
\resizebox{1.0\textwidth}{!}{%
\begin{minipage}{\textwidth}
\textbf{Input:}\\
$\left\{ B_{k} \right\}$: Bone vectors from SMPL joints\\
$\left\{ O^s \right\}$: Open contours on a source mesh $S$\\
$\left\{ O^t \right\} $: Open contours on a target mesh $T$\\
\textbf{Output:}\\
$\left\{O^s,O^t,d^{st}\right\}$: A set of open contour pairs ($O^s \in S$, $O^t \in T$) and their distance along the skeleton\\
\renewcommand{\algorithmicrequire}{\textbf{Procedure:}}
\renewcommand{\algorithmicensure}{\textbf{return:}}
\newcommand{\INDSTATE}[1][1]{\STATE\hspace{#1\algorithmicindent}}
\caption{Axial distance computation among open contours.}
\label{alg:lg}
\begin{algorithmic}
\State
\Require MeasureDistanceAlongBones($\left\{ B_{k} \right\}$, $\left\{ O^s \right\}$, $\left\{ O^t \right\} $)
      \For {\textbf{each} $O^s$}
          \State $ \left\{L_{i}^s\right\}  \gets $EncircledBones $(O^s,\left\{ B_{k} \right\}) $
    \EndFor

    \For {\textbf{each} $O^t$}
          \State $ \left\{L_{j}^t\right\}  \gets$ EncircledBones $(O^t,\left\{ B_{k} \right\}) $
    \EndFor
    \State  $MAP \gets \emptyset$
      \For {\textbf{each} $(O^s,O^t)$ pair}
      \State $MAP^{st} \gets \emptyset$
      \For {\textbf{each} $(L_i ^s, L_j^t)$ pair}
            \If {bodypartname($B_{i}^s$) == bodypartname($B_{j}^t)$}     \ \  //bodyparts: torso and four limbs
            \State $MAP^{st} \gets MAP^{st} \cup \{ (O^s,O^t,L_{i}^s, L_{j}^t) \} $  
            \EndIf
       \EndFor
       \State $MAP \gets MAP \cup \left\{MAP^{st}\right\}$
      \EndFor

      \For {\textbf{each} $MAP^{st}$ \textbf{in} $MAP$}   
      \State 1. take $\left \{ (L_{i}^s, L_{j}^t) \right \}  $ from $ MAP^{st}$
      \State 2. $i^* \gets \mathop{\arg\min}_{i} \| h_i^s \in L_i^s\|$
      \State 3. take $P_{i^*}^s, P_{j^*}^t$ from $(L_{i^*}^s, L_{j^*}^t)$
      \State 4. $d(O^s,O^t) \gets distToRoot(P_{j^*}^t)-distToRoot(P_{i^*}^s)$
      \EndFor
\Ensure  $\left\{O^s, O^t, d^{st}=d(O^s,O^t)\right\}$
\State
\Require EncircledBones($O,\left\{ B_{k} \right\}$)\\
\textbf{Input:}\\
$\left\{ B_{k} \right\}$: bone vectors from SMPL joints\\
$O$: open contour on a mesh\\
\textbf{Output:}\\
$ L= \{ L_i \}$, $L_i=\{(B_{i}, u_{i}, P_{i}, h_{i})\}$ where $B_{i}$ is bone vector enclosed by $O$, and $u_{i}$ and $h_{i}$ are the axial and the perpendicular distances of the contour center in relation to $B_{i}$. 
\\
      \State $L \gets \emptyset, O_{center} \gets center(O)$
      \For {\textbf{each} $B_{i}$ \textbf{in} $\left\{ B_{k} \right\}$}
          \State 1. $J_{i}^{parent}, J_{i}^{child} \gets \text{Two joints of bone vector }B_{i}$ 
          \State 2. $u_{i} \gets \frac{\overrightarrow{J_{i}^{parent}J_{i}^{child}}\cdot\overrightarrow{J_{i}^{parent}O_{center}}}{\|\overrightarrow{J_{i}^{parent}J_{i}^{child}} \|^2}$  
          \hspace{0.7cm} \hspace{1.4cm} // axial distance from the parent
          \State 3. $P_{i} \gets J_{i}^{parent}+u*(\overrightarrow{J_{i}^{parent}J_{i}^{child}}))$  \hspace{1.2cm} // projection of $O_{center}$ onto the bone
          \State 4. $h_{i} \gets \|O_{center}- P_{i}\|$  \hspace{3cm}// distance to the bone 
          \If {$0<u_i<1$ }
          \State $L_i \gets \left\{B_{i}, u_{i},P_{i},h_{i}\right\}$
          \State $ L \gets L \cup L_i $
      \EndIf
      \EndFor
\Ensure $ L= \left \{ L_i \right \} $
\end{algorithmic}
\end{minipage}%
}
\end{algorithm*}

\end{document}